\documentstyle[prb,aps,psfig,multicol]{revtex}
\begin{document}
\draft
\title{Boundary and Bulk Phase Transitions in the Two Dimensional $Q>4$ 
State Potts Model}
\author{Ferenc Igl\'oi}
\address{Research Institute for Solid State Physics and Optics, 
H-1525 Budapest, P.O.Box 49, Hungary\\
Institute for Theoretical Physics, Szeged University, 
H-6720 Szeged, Hungary}
\author{Enrico Carlon}
\address{Institute for Theoretical Physics, Katholieke Universiteit Leuven,
Celestijnenlaan 200D, B-3001 Leuven, Belgium}

\date{Preprint KUL-TF-98/22}

\maketitle

\begin{abstract}
The surface and bulk properties of the two-dimensional $Q>4$ state 
Potts model in the vicinity of the first order bulk transition point 
have been studied by exact calculations and by density matrix 
renormalization group techniques. For the surface transition the 
complete analytical solution of the problem is presented 
in the $Q \to \infty$ limit, including the critical and tricritical 
exponents, magnetization profiles and scaling functions. We have shown 
in an accurate numerical study that the universality class of the 
surface transition is independent of the value of $Q>4$. For the bulk 
transition we have numerically calculated the latent heat and the 
magnetization discontinuity and we have shown that the correlation 
lengths in the ordered and disordered phases are identical at the 
transition point.
\end{abstract}


\newcommand{\bc}{\begin{center}}
\newcommand{\ec}{\end{center}}
\newcommand{\be}{\begin{equation}}
\newcommand{\ee}{\end{equation}}
\newcommand{\beqn}{\begin{eqnarray}}
\newcommand{\eeqn}{\end{eqnarray}}

\begin{multicols}{2} \narrowtext
\section{Introduction}
\label{sec:intr}

The two-dimensional $Q$-state Potts model \cite{wu} is an important 
testing ground of statistical physics for different approximate theories 
and numerical methods. For $Q \le 4$ the model has a second order 
phase transition and the critical exponents, both in the bulk and at 
the surface, are $Q$-dependent as they have been predicted by conformal 
invariance \cite{cardy} and by Coulomb-gas methods \cite{nienhuis}.

On the other hand in the first order regime (i.e. $Q > 4$) there are
analytical results \cite{baxterbook} about the properties of the 
system {\it at the transition point}, such as the latent heat, 
$\Delta E$, the discontinuity of the magnetization \cite{baxtermagn},
$\Delta m$, and the correlation length in the disordered phase \cite{ksz}, 
$\xi_d$, but still some issues are unsolved.
One of these is the relation between the correlation lengths in the order
($\xi_o$) and in the disordered ($\xi_d$) phase for which different 
theoretical approaches which predict either \cite{janke1} $\xi_o/\xi_d=1/2$
or $\xi_o/\xi_d=1$, where the latter is supported by recent Monte Carlo 
simulations \cite{janke}. 

Another unclarified problem of the $Q>4$ model concerns the behavior of 
the surface phase transition. 
In studies of surface critical behavior \cite{binder} most of the theoretical 
investigations have been focused on the so-called ordinary transition 
which occurs at a bulk continuous transition in absence of enhanced 
surface couplings and/or surface fields. In this case scaling ideas 
and, in two dimensions, conformal invariance \cite{cardy} are often 
used to predict surface critical exponents and scaling functions.

A completely different approach is necessary for systems with a 
discontinuous bulk phase transition, like the $Q>4$ state Potts 
model, where scaling and conformal invariance do not hold.
It was Lipowsky \cite{lipowskyprl} who first showed, in a mean-field 
calculation, that one can have a first order transition in the bulk 
and a second order one at the surface. This phenomenon, known as 
{\it surface induced disorder} (SID) and closely related to the 
wetting phenomena \cite{dietrich}, has been studied at the mean 
field level \cite{lipowskyreview,krollmf}, by means of Monte Carlo 
simulations \cite{mcsid} and in some experimental systems \cite{dosch}, 
but no exact results are available on {\it microscopic models} so far.

In this paper we present exact and numerical results on surface 
phase transitions in the two-dimensional $Q>4$ state Potts model. 
Our study is based 
on two complementary approaches. First, we show that there exists 
a systematic large $Q$ expansion of the problem, which is exactly 
soluble in the $Q \to \infty$ limit. We show that in this limit 
there is SID with a surface magnetization which vanishes continuously 
by approaching the bulk transition temperature and we find the exact values 
of the whole sets of surface exponents. At a sufficiently large value 
of a surface field applied to the spins at the free boundary, the 
surface transition becomes first order.
The continuous and first order regimes are separated by a tricritical 
transition whose exponents are also found exactly.

The analytical results are complemented by numerical investigations 
based on density matrix renormalization group (DMRG) techniques 
\cite{whitePRL}, through which we calculate magnetization and energy 
density profiles in $L \times \infty$ strips for $5 \le Q \le 9$. 
Following the approach of Ref. \onlinecite{carlonigloi} we restricted 
ourselves to moderately large values of $L$; typically we considered 
$L \leq 40$ and occasionally also larger systems (up to $L = 80$). 
We analyzed the numerically accurate data by efficient extrapolation 
techniques and using information of scaling functions obtained in the 
exact solution at $Q \to \infty$. We found again SID and strong indications 
that the surface exponents do not depend on the value of $Q$.

The same type of DMRG technique is used also to calculate properties of 
the bulk transition. Here we note that for weak first-order transitions, 
when $Q$ is close to four, it is a challenge for numerical methods 
\cite{pottsq5} (Monte Carlo simulation, renormalization group (RG) methods, 
finite-size scaling, series expansion, etc.) to detect the correct order 
of the transition and to estimate the corresponding discontinuities. 
In all cases studied we found good agreement between the numerically 
extrapolated values and the exact results, which once again proves 
the powerfulness and reliability of the DMRG algorithm. The numerical
calculation of the bulk correlation length in the ordered phase 
($\xi_d$) for $Q=9$ confirms with good accuracy the conjectured relation 
$\xi_d = \xi_o$. 

The paper is organized as follows. In Section \ref{sec:form} we introduce 
the model, define the basic quantities and outline the use of the DMRG 
method for the problem. In Section \ref{sec:bulk} we present some numerical 
results concerning bulk quantities and analyze the behavior of the bulk 
correlation length. The surface transition of the model is studied in 
Section \ref{sec:surf} by analytical and numerical methods. 
Section \ref{sec:disc} concludes the paper with some discussion.

\section{Formalism}
\label{sec:form}

We consider the $Q$-state Potts model in a $L \times \infty$ square lattice 
defined by the Hamiltonian
\beqn
-\beta H=K_1\sum_{k=-\infty}^{\infty}\sum_{l=1}^{L-1} \delta(s_{l,k}-s_{l+1,k})
\nonumber\\
+ K_2\sum_{k=-\infty}^{\infty}\sum_{l=1}^{L} \delta(s_{l,k}-s_{l,k+1})\;,
\label{clhamilton}
\eeqn
in terms of a $Q$ component spin variable $s_{l,k}=1,2,\dots,Q$. We use for
the inverse temperature $\beta=1$ and consider the ferromagnetic model
$K_1,K_2 > 0$, where the transition temperature is given from a duality
relation as:
\be
\left(e^{K_1}-1\right) \left(e^{K_2}-1\right)=Q\;.
\label{crtemp}
\ee
We work in the transfer matrix formalism, with the transfer matrix along the
strip given by \cite{mittagstephen}:
\beqn
{\cal T}=\exp\left[{K_1^* \over Q} \sum_{l=1}^L\sum_{q=1}^{Q-1} M_l^q \right]
\nonumber\\ \times
\exp\left[K_2 \sum_{l=1}^{L-1}\left(\delta(s_l-s_{l+1})- {1 \over Q} \right)
\right]\;,
\label{trmatrix}
\eeqn
where $M_l$ is a 
spin-flip operator: $M_l| s_l \rangle=| s_l +1,{\rm mod}~Q \rangle$ 
and $K_1^*$ is the dual coupling defined as $(e^{ K_1} -1)(e^{K_1^*} 
-1)=Q$, thus at the critical point $K_1^*=K_2$.

Different physical quantities of the model can be expressed by the 
leading and next to leading eigenvalues 
of ${\cal T}$, 
$\Lambda_0$ and $\Lambda_1$, respectively, and by the corresponding 
dominant eigenvector $|v_0\rangle$. For example one obtains for the 
magnetization profile 
\begin{eqnarray}
m_l = \frac{Q \, \langle v_0| \delta(s_l-1) |v_0 \rangle -1}{Q-1},
\label{magnet}
\end{eqnarray}
and for the energy density
\be
\epsilon_l=\langle v_0| \delta(s_{l,k}-s_{l+1,k}) |v_0 \rangle \;.
\label{ener}
\ee
The correlation length along the strip is given by
\be
\xi_{\parallel}=\log^{-1} (\Lambda_0 /\Lambda_1)\;.
\label{corrlength}
\ee
In what follows we use symmetry breaking boundary conditions (BC)
and fix spins along one of the surfaces to the same state $s_{1,k}
=1,~~k=0,\pm1,\dots$ and at the other surface of the strip we apply 
a surface field of the form $-H_s \sum_k \left[\delta(s_{L,k}-1)-1
/Q\right]$.

In the transfer matrix approach one needs to diagonalize the transfer 
matrix, which is technically very demanding, since the dimension of 
the matrix being $Q^{L-1} \times Q^{L-1}$ grows very rapidly with the 
width of the strip. With the traditional finite size scaling methods 
based on the use of the L\'anczos algorithm one can only treat very 
limited range of the values of $L$, especially for not too small $Q$-s. 
This limitation is the reason why density profiles and surface quantities 
have not been studied following this method.
Another possibility is the representation of the Potts variables in 
terms of Temperly-Lieb (TL) operators. This has the advantage that 
the transfer matrix dimension does not depend on Q, which enters in
the calculation as a parameter, therefore this representation could 
be suitable for studying large Q. However quantities as density profiles 
cannot be directly expressed by the TL operators. For the study of these
quantities one should then use the DMRG in the original Potts representation.

In a previous paper\cite{carlonigloi}, hereafter referred to as Paper I, we 
have used the DMRG method for the $Q \le 4$ Potts models to study the critical
properties, such as density profiles, Casimir amplitudes and critical exponents. 
Here we extend the investigations to the first order transition regime with $Q>4$.

\begin{figure}[b]
\centerline{\psfig{file=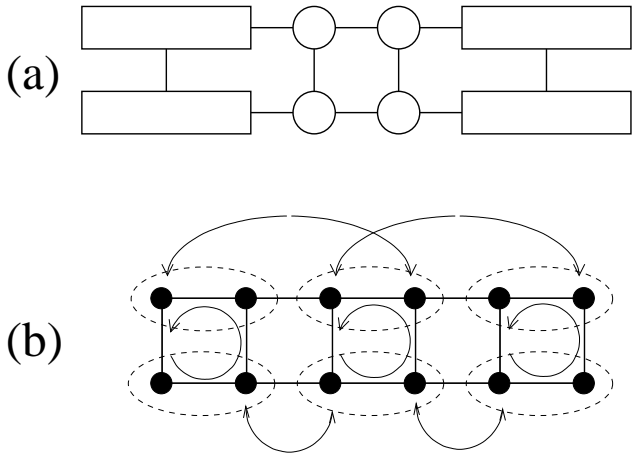,height=6cm}}
\vspace{5mm}
\caption{(a) Schematic view of a transfer matrix element as 
generated by the DMRG algorithm; the blocks describe 
approximately a large number of spins using $m$ states only. 
(b) Pseudospin version of the $Q=q^2$ Potts model as 
described in the text; two $q$ spins grouped by the dashed 
ellipses form a $Q$ spin and interactions are defined on
plaquettes and between further neighbors along the row as
indicated in the figure.}
\label{FIG01}
\end{figure}

The principle and the technical details of the DMRG method are well
documented in the literature \cite{whitePRB} and its application to 
the $Q$-state Potts model is described in I. 
Here we just mention that in the DMRG method a strip of large
width $L$ is described by an approximated transfer matrix as 
represented in Fig. \ref{FIG01}(a), where the rectangles 
denotes blocks containing a large number of spins, but described 
by only $m$ states. For the Potts model the total dimension of 
the matrix would be $m^2 Q^2 \times m^2 Q^2$. Although $m$, the 
number of states kept, can be small (typically $m \sim 100$) the 
algorithm provides accurate results even for large $L$. For a 
strip of given width, as $Q$ increases, one needs to increase 
$m$ in order to have sufficient accuracy. This limits the range 
of values of $Q$ that can be investigated with this method.
Therefore for the $Q=9$ model
we have implemented a method, which in principle can be used 
for $Q=2^2,~3^2,~4^2,\dots$, and which considerably speeds up
the numerical computation. 
It consists in rewriting the original Hamiltonian for the $Q = 
q^2$-Potts model as a $q$-Potts model with plaquette and further 
neighbor interactions as illustrated in Fig. \ref{FIG01}(b). 
Two neighboring $q$-spins (or {\em pseudospins}) in a row form a 
$Q$-spin; in terms of these new variables which we indicate as
$\tilde{s}_{i,j}$ the Hamiltonian (\ref{clhamilton}) takes the
following form:
\beqn
-\beta H &=& K_1\sum_{k=-\infty}^{\infty}\sum_{l=1}^{L-1} 
\delta(\tilde{s}_{2l-1,k} - \tilde{s}_{2l+2,k})
\delta(\tilde{s}_{2l,k} - \tilde{s}_{2l+1,k})
\nonumber \\
&+& K_2\sum_{k=-\infty}^{\infty}\sum_{l=1}^{L} 
\delta( \tilde{s}_{l,k}-\tilde{s}_{l,k+1})
\delta( \tilde{s}_{2l,k}-\tilde{s}_{2l,k+1})\;.
\label{clhamilton2}
\eeqn

When the DMRG algorithm is applied to this Hamiltonian one 
deals with matrices of dimension $m^2 Q \times m^2 Q$, and
the number of states kept in the blocks can be enlarged, with 
respects of the original representation. Comparisons of the 
standard and pseudo-spin version for the $Q = 9$ case indicate 
that the latter provides a remarkable increase in computational 
speed and accuracy.
We also point out that a similar idea has been applied recently
to study of quantum systems with a large number of states per site
\cite{erik}.

Before closing this section we turn back to the transfer matrix 
in (\ref{trmatrix}), which in the vicinity of the critical point
can be simplified in the {\it anisotropic} or Hamiltonian limit 
of strong vertical and weak horizontal couplings where both 
$K_1^*$ and $K_2$ are small.
Then one can combine the exponentials to obtain ${\cal T} =
\exp(-K_2 {\cal H})$ with a Hamilton operator\cite{solyompfeuty}
\be
{\cal H}=-\sum_{l=1}^{L-1} \delta(s_l-s_{l+1}) - h \sum_{l=1}^L 
\sum_{q=1} ^{Q-1} M_l^q\;,
\label{hamilton}
\ee
up to a constant term and $h=K_1^*/K_2 Q$. Thus at the critical 
point $h_c=1/Q$.  

In this Hamiltonian version of the model quantities of physical 
interest are derived from the ground state wavefunction $|\Psi_0 
\rangle$ and from the ground state and first excited state energies 
$E_0$ and $E_1$. In the anisotropic limit the transfer matrix and the 
Hamilton operator commute, $[{\cal T},{\cal H}]=0$; the corresponding 
eigenvectors are the same, $|\Psi_0 \rangle=|v_0 \rangle$, consequently 
the density profiles in Eqs.(\ref{magnet}) and (\ref{ener}) stay in the 
same form. On the other hand the correlation length along the strip is 
now expressed by the inverse gap
\be
\xi_{\parallel}=1/(E_1-E_0)\;,
\label{hamcorr}
\ee
instead of Eq.(\ref{corrlength}). In the Hamiltonian version the symmetry 
breaking BC are equivalent to fix one of the surface spins to the state 
$|s_1\rangle= |1\rangle$ and to add a term $-H_s [ \delta(s_L-1)-1/Q ]$ to 
${\cal H}$ in (\ref{hamilton}) in order to study the effect of a surface 
field $H_s$ on the system. This Hamiltonian version of the model will be 
used to solve the surface properties of the model in the $Q \to \infty$ limit.

\section{Bulk transition}
\label{sec:bulk}

In this Section we consider the isotropic model, $K_1=K_2$, and present
numerical results about the bulk transition obtained by the DMRG method
for $5 \le Q \le 9$. Our aim in calculating the latent heat and the
magnetization discontinuity is to demonstrate the accuracy of the DMRG
approach. On the other hand we used the numerical results about the
correlation length in the ordered phase at the transition point to
analyze the ratio $\xi_o/\xi_d$.

We start with the calculation of the latent heat, $\Delta E$, which 
is just the difference in the energy densities in the ordered and the 
disordered phases at the transition point $\Delta E=\epsilon^o-\epsilon^d$.
The arithmetic average of the above two energy densities is the
energy per bond, $\varepsilon=(\epsilon^o+\epsilon^d)/2$, which is known 
from self-duality at the transition point \cite{wu}:
\be
\varepsilon={1 \over 2} \left( 1 + {1 \over \sqrt{Q}} \right)\;.
\ee
Consequently it is enough to calculate the energy density in the ordered
phase and the latent heat then follows as:
\be
\Delta E=2(\epsilon^o - \varepsilon)\;.
\ee
To estimate $\epsilon^o$ we have considered $L \times \infty$ systems 
with fixed spin BC at both edge of the strip and calculated by the DMRG 
method the energy density in (\ref{ener}) at the middle of the strip, 
$\epsilon_{L/2}$, for a series of the lengths $L=8,12,16,\dots,L_{max}$. 
We selected moderately large finite systems, $L_{max}\sim 40-80$,
so that the DMRG data are still accurate enough (with a relative
accuracy of $10^{-5}-10^{-6}$) to use sophisticated series extrapolation
procedures, such as the BST algorithm \cite{BST}. As we have shown in 
paper I, this type of combination of the DMRG method and the finite-size 
extrapolation methods results in accurate data in the thermodynamic 
limit. (In a very recent corner transfer matrix DMRG calculation
of the latent heat of the $Q=5$ Potts model Nishino and Okunishi\cite{Q5}
used another strategy and performed the calculation for several thousand 
of layers without using series extrapolation algorithms.)
To demonstrate the accuracy of our approach we show in Table \ref{TAB01} 
the BST extrapolants for the latent heat of the $Q=8$ model. As seen in 
this Table the extrapolants seem to converge to a value of $\Delta 
E(Q=8)=0.242(2)$, which is fairly close to the exact value $0.2432$. 
Results of the extrapolated latent heats for $Q=5,6,8$ and $9$, together 
with the exact values are presented in Table \ref{TAB02}. It is seen that 
the DMRG method combined with series extrapolation techniques predicts 
the right order of the transition and the estimated latent heats are in 
good agreement with the exact results, even for $Q=5$. As expected, the 
accuracy of the numerical results is lower for weakly first order 
transitions, i.e. which are characterized by a small latent heat.

Next we turn to calculate the magnetization discontinuity, $\Delta m$, 
at the transition point. Since the magnetization in the disordered 
phase is zero $\Delta m$ is just the bulk magnetization in the ordered 
phase. This can be estimated from the magnetization profile in 
(\ref{magnet}) in the middle of the strip, $m_{L/2}$, with fixed spin 
BC at both edges of the strip. As for the latent heat we have calculated 
$m_{L/2}$ for a series of moderately large $L$-s by the DMRG method and 
then performed an extrapolation with the BST procedure. The estimated 
values of $\Delta m$ and the corresponding exact values are collected 
in Table \ref{TAB02} for different values of $Q$.
As seen in the Table the DMRG results agree well with the exact results 
and the relative accuracy in this case is even higher than for the latent
heat. This is due to the fact that the magnetization discontinuity is 
relatively large, even for $Q=5~~\Delta m$ is about $50 \%$ 
of the saturation value.
Considering the finite size corrections it is interesting to note that
while these corrections for $Q=5$ are approximately logarithmic, for 
$Q=9$ we have a $L^{-2}$ correction behavior.

Finally, at the bulk transition point we have calculated the correlation
length in the ordered phase, $\xi_o$, 
for $Q=9$, since for smaller $Q$-s the correlation length 
exceeds the available size of the strip. 

\begin{figure}[b]
\centerline{\psfig{file=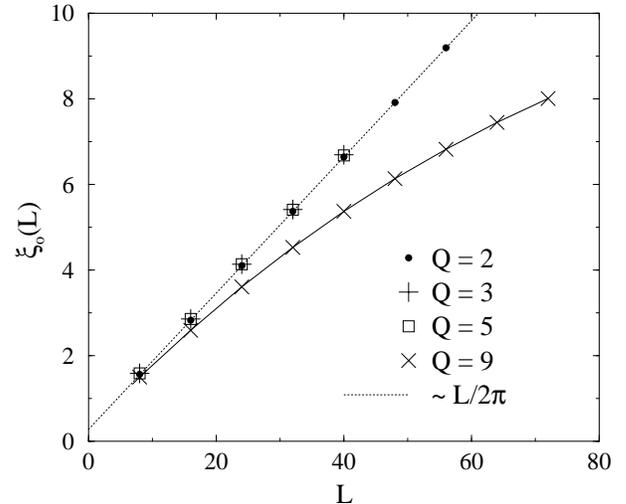,height=7cm}}
\vspace{5mm}
\caption{Correlation length $\xi_o (L)$ for fixed/fixed boundary 
conditions as function of the strip width $L$. The dotted line is the
conformal result.}
\label{FIG02}
\end{figure}

Figure \ref{FIG02} shows a plot of the correlation length $\xi_o (L)$ for 
$Q = 2$, $3$, $5$ and $9$, calculated from the ratio of the largest and the 
second largest eigenvalues of the transfer matrix in Eq. (\ref{corrlength}).
The $Q = 2$ and $3$ cases follow the expected behavior derived from conformal 
invariance \cite{cardy} for a strip with fixed and equal spins at the boundary, 
i.e. $\xi_o (L) \simeq L/(2 \pi)$. 
Notice that also in the $Q = 5$ case $\xi_o (L)$ follows a straight line with 
the same slope as for the $Q = 2$ and $3$. The transition, as pointed out above, 
is only weakly first order in the case $Q = 5$ with a correlation length much 
larger than the largest system size available here. The data for $Q = 9$ show a 
clear sign of saturation for large $L$.
The extrapolation to $L \to \infty$ has been performed by the BST method and 
yields $\xi_o = 15.0(4)$. This quantity is in good agreement with the exactly 
known correlation length of the disordered phase which is $\xi_d = 14.9$\cite{ksz}.
These results corroborates the recent conjecture of Ref. \onlinecite{janke},
namely that $\xi_o = \xi_d$, based on results from Monte Carlo simulations.

\section{Surface transition}
\label{sec:surf}

The surface transition of the model is studied by two complementary methods.
First, we present the analytical solution of the problem in the $Q \to \infty$
limit, then the transition for finite $Q$-s are studied numerically by the
DMRG method. Also in this numerical part of the work we use the analytical
results obtained for $Q \to \infty$ about the critical singularities and 
scaling functions, as a help in analyzing the DMRG data.

\subsection{Analytical solution in the $Q \to \infty$ limit}

Here we work with the anisotropic version of the model, which 
is characterised by the Hamiltonian in (\ref{hamilton}) and use 
the BC described below (\ref{hamilton}), i.e. fixed spin at $i=1$ 
and a surface field of strength $H_s$ in the other end, $i=L$. 
(For a free surface we have $H_s=0$.)

The Hamiltonian ${\cal H}$ is represented by a matrix of dimensions 
$Q^{L-1} \times Q^{L-1}$, which can be further reduced in a 
block-diagonal form by making use of the symmetries of the problem. 
The dimension of the block which contains the ground state (and also 
the first excited state) is independent of the value of $Q$\cite{igloisolyom}. 
For small systems, $L=2$ and $L=3$, this block, which has the 
highest symmetry, has been explicitly constructed in the Appendix. 
Analyzing the structure of this ground-state block one can notice 
further simplifications in the vicinity of the critical point, where 
one can perform a large $Q$ expansion, such that the corrections to 
the leading behaviour are of $O(Q^{-1/2})$. 
As shown in the Appendix in the limit $Q \to \infty$ the ground-state 
block has a dimension of $L$ with an eigenvalue matrix ${\tilde H}$ 
given by
\be
{\tilde H} =-(L-1)-h\sqrt{Q} \left(
\matrix{
h_s &  1  &      &       &     0   \cr
  1 &  t  &   1  &       &         \cr
    &  1  & 2t   &\ddots &         \cr
    &     &\ddots&\ddots &      1  \cr
  0 &     &      &   1   &(L-1)t   \cr}
\right)\;.
\label{htilde}
\ee
Here $t = (h Q - 1)/ h \sqrt{Q}$ measures the distance from the bulk transition point 
and $h_s = H_s/h \sqrt{Q}$. Denoting the components of the eigenvectors of 
${\tilde H}$ as $\psi_i(n)$ with $n = 1$, \ldots, $L$ one can express the 
magnetization profile in terms of the ground state wavefunction as follows:
\be
m_l=\sum_{n=1}^{L+1-l}\left[\psi_0(n)\right]^2\;.
\label{magnet1}
\ee

{\it At the transition point} $t=0$ the matrix (\ref{htilde}) is exactly 
diagonalizable with eigenvectors
\be
\psi_i(n)=A_i \sin [ k_i (L+1-n)] ,
\label{soleigvec}
\ee
and eigenvalues
\be
E_i=-(L-1)-2h\sqrt{Q}\cos k_i ,
\label{soleigval}
\ee
where $i=0,1$, \ldots, $L-1$ and the wavenumbers $k_i$ are 
determined by the equation:
\be
h_s \sin (L k_i) =\sin [(L+1) k_i ]\;.
\label{ki}
\ee
As known by standard analysis the spectrum and the ground state of ${\tilde H}$ 
have different properties for $h_s<1$ and $h_s>1$, respectively. In the
following we treat the different regions of surface transition separately.

\subsubsection{Critical surface transition regime - $h_s < 1$}

For $h_s<1$ there are $L$ real solutions of Eq.(\ref{ki}) and the spectrum of
${\tilde H}$ is quasi-continuous. The qualitative behaviour of the surface
transition is the same in the whole region, therefore we consider here the
free surface case, $h_s=0$, when the the solution of (\ref{ki}) is given
explicitly as:
\be
k_i={\pi (i+1) \over L+1}~~,~~h_s=0\;,
\label{hs0}
\ee
and the normalization in (\ref{soleigvec}) is $A_i=[2/(L+1)]^{1/2}$.

The boundary magnetization from (\ref{magnet1}) is given by:
\be
m_s=m_L={2 \over L+1} \sin^2\left( {\pi \over L+1} \right) \sim L^{-3}\;,
\label{msh0}
\ee
thus it goes zero at the bulk transition point, $t=0$. Consequently the
surface phase transition is of {\it second order}, furthermore the
anomalous dimension of the surface magnetization operator is
\be
x_s^{\rm cr}=3\;,
\label{xscr}
\ee
as obtained from the finite size dependence in (\ref{msh0}).

\begin{figure}[b]
\centerline{\psfig{file=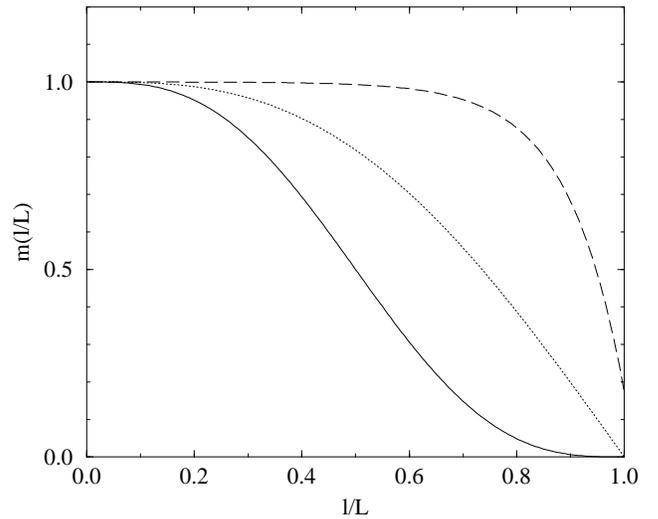,height=7cm}}
\vspace{5mm}
\caption{Magnetization profiles of the $Q \to \infty$ Potts model 
at the bulk transition point with fixed spin BC at $l=1$ and with 
different values of the surface field, $h_s$, at $l=L$. Solid line 
- critical surface transition, $h_s=0$; dotted line - tricritical 
surface transition, $h_s=1$;
dashed line - first order surface transition, $h_s > 1$.}
\label{FIG03}
\end{figure}

The magnetization profile from (\ref{magnet}) can also be evaluated, which
in the continuum approximation $L,l \gg 1$ and as a function of the distance
from the free surface, $\Delta l=L+1-l$, is given by
\beqn
m_{\Delta l}={\Delta l \over L} - {1 \over 2\pi}
\sin \left( 2\pi {\Delta l \over L} \right)
\sim \left({ \Delta l \over L}\right)^3+
O\left[\left({ \Delta l \over L}\right)^5\right],
~~~~~~h_s=0\;.
\eeqn
This profile is shown as a solid line in Fig. \ref{FIG03}. 

In a strip of width $L$ the correlation length parallel with the free
surface is given from (\ref{hamcorr}):
\be
\xi_{\parallel}^{-1}=2h\sqrt{Q} \left[ \cos \left({\pi \over L+1}\right)-
\cos \left({ 2 \pi \over L+1}\right) \right] \sim L^{-2}\;.
\label{corrh0}
\ee
Since the correlation length perpendicular to the strip is just $\xi_{\perp}
\sim L$, we have the asymptotic relation $\xi_{\parallel} \sim \xi_{\perp}^
{z^{\rm cr}}$, with an anisotropy exponent
\be
z^{\rm cr}=2\;.
\label{zcr}
\ee
Away from the bulk transition point for $0 < -t \ll 1$ the boundary magnetization
can be studied by a perturbation expansion, when the diagonal term of
$\tilde{H}$ in (\ref{hamilton}), $\tilde{H}_{i,j}=-t(i-1)\delta_{i,j}$, is considered
as a perturbation. From this calculation the first order correction to the
boundary magnetization is given for $L \to \infty$ and $0<-t \ll 1$ as:
\be
m_s={2 \pi^2 \over L^3} + {1 \over 2} \sqrt{Q} t + \dots\;
\ee
Thus the surface magnetization exponent, defined in the thermodynamic limit
($L \to \infty$) as $m_s \sim (-t)^{\beta_s^{\rm cr}}$ is given by:
\be
\beta_s^{\rm cr}=1\;.
\ee
The small $t$-behavior of the system can also be investigated in 
the frame of a continuum approximation, when the eigenvalue problem 
of $\tilde{H}$ in
(\ref{htilde}),
which is a second-order difference equation, is transformed
into a differential equation of the continuous variable $x=i$:
\be
{\partial^2 \psi \over \partial x^2} + (\epsilon+\tilde{t}x)\psi(x)=0\;,
\label{diffeq}
\ee
with $\epsilon=2+E/(h\sqrt{Q})$ and $\tilde{t}=t/(h \sqrt{Q})$, whereas the
boundary conditions are $\psi(x=0)=\psi(x=L)=0$. From the properties of the
solution of (\ref{diffeq}) one obtains for the temperature dependence of the
correlation lengths as:
\be
\xi_{\parallel} \sim |t|^{-2/3}~~,~~\xi_{\perp} \sim |t|^{-1/3}\;,
\label{tcorr}
\ee
thus the corresponding critical exponents are $\nu_{\parallel}^{\rm cr}
=2/3$ and $\nu_{\perp}^{\rm cr}=1/3$ and the scaling relation 
$\nu_{\parallel}/\nu_{\perp}=z^{\rm cr}$ is satisfied with $z^{\rm cr}$ 
in (\ref{zcr}). The exponents of the surface critical transition 
are collected in Table \ref{TAB03}.

\subsubsection{First-order surface transition regime - $h_s>1$}

The qualitative behavior of the solution of $\tilde{H}$ is changing when
the strength of the surface field, $h_s$ grows over $h_s=1$. For $h_s>1$
the ground state of $\tilde{H}$ has the form of a localised mode and the
corresponding wavenumber in (\ref{ki}) is complex, $k_0=i \kappa$. In the
limit $L \to \infty$ the wavefunction of this localised mode is given by
\be
\psi^{loc}(n)=(h_s^2-1)^{1/2} h_s^{-n}\;,
\label{psiloc}
\ee
with the energy
\be
E^{loc}=-(L-1)-h\sqrt{Q}(h_s+h_s^{-1})\;,
\label{Eloc}
\ee
which is separated from the continuum part of the spectrum given in the
form of (\ref{soleigval}).

The surface magnetization at the bulk transition point, $t=0$, from
(\ref{magnet1})
\be
m_s=1-{1 \over h_s^2}\;
\ee
is finite, thus the surface transition is of {\it first-order} for 
$h_s>1$. The $\delta_{1,1}$ exponent defined as $m_s \sim (\delta 
h_s)^{1/\delta_{1,1}}$ with $\delta h_s=h_s-1$ is given by:
\be
\delta_{1,1}^{\rm tr}=1\;,
\ee
where the superscript ``tr" refers to the tricritical point, which is
approached as $h_s \to 1^+$.

The surface correlation length parallel with the surface is finite in the
$L \to \infty$ limit
\be
\xi_{\parallel}^{-1}=E_1-E^{loc}=h\sqrt{Q} (h_s+h_s^{-1}-2) \sim (\delta h_s)^2\;,
\label{xi1}
\ee
as expected for a discontinuous surface transition. In (\ref{xi1}) for $E_1$
the edge of the continuum spectrum in (\ref{soleigval}) is used. The
magnetization profile as a function of the distance from the free surface,
$\Delta l=L+1-l$, is given by: 
\be
m_{\Delta l}=1-h_s^{-2\Delta l}~~~~~~~~~~~~~~~h_s > 1,~L \to \infty\;.
\ee
As seen from the dashed line of Fig. \ref{FIG03} there is a 
surface region of the profile having a size of
\be
\xi_{\perp}(h_s)={1 \over 2 \ln h_s} \sim (\delta h_s)^{-1}\;,
\label{xiperp}
\ee
which defines the perpendicular correlation length at the surface. As
the tricritical point is approached, $h_s \to 1^+$, the two correlation
lengths are related as $\xi_{\perp}^2 \sim \xi_{\parallel}$, thus the
anisotropy exponent in the tricritical point
\be
z^{tr}=2\;,
\label{ztr}
\ee
is the same as in the critical region Eq.(\ref{zcr}).

\subsubsection{Tricritical surface transition - $h_s=1$}

As we mentioned before the tricritical point with $h_s=1$ separates the
continuous and first-order surface transition regimes. At the bulk
transition point, $t=0$, the solution of (\ref{ki}) for the wavenumbers
is given by
\be
k_i={\pi \over 2L+1}(2i+1)~~,~~h_s=1\;,
\label{ki1}
\ee
and the normalization in (\ref{soleigvec}) is $A_i=2/\sqrt{2L+1}$.

The surface magnetization
\be
m_s={4 \over 2L+1} \cos^2\left({\pi \over 2(2L+1)}\right) \sim L^{-1}\;,
\ee
goes to zero as $L \to \infty$, however the tricritical surface magnetization
scaling dimension
\be
x_s^{\rm tr}=1\;,
\ee
is different from its value in the critical region (\ref{xscr}).

The tricritical magnetization profile can be evaluated from 
(\ref{magnet}), which in the continuum approximation is given by
\beqn
m_{\Delta l} &=&{\Delta l \over L} + {1 \over \pi}
\sin\left( \pi {\Delta l \over L} \right)\sim 
\nonumber \\
& \sim &
{\Delta l \over L}+O\left[\left({\Delta l \over L} 
\right)^3\right],~~~~~~~~~h_s=1\;.
\eeqn
as a function of the distance from the free surface, $\Delta l$.
This profile is also shown in Fig. \ref{FIG03} as a
dotted line where one can notice 
the
different 
behavior close to the free surface for the critical and tricritical 
cases. This is due to the fact that the profiles grow with the 
respective $x_s$ scaling dimensions, which are different in the 
two fixed points.
The correlation length parallel with the surface can be calculated as in
(\ref{corrh0}) with the asymptotic behavior $\xi_{\parallel}=L^2$. Thus
the anisotropy exponent is $z^{\rm tr}=2$ in accordance with the previous
derivation in (\ref{ztr}).

Away from the bulk transition point for $|t| \ll 1$ one can repeat the
calculations as described for the critical surface transition. In the
continuum approximation the differential equation in (\ref{diffeq})
remains the same and just the boundary condition at $x=0$ will be 
changed to $\partial \psi / \partial x|_{x=0}=0$. The tricritical 
exponents obtained from the solution of the differential equation 
are presented in Table \ref{TAB03}, together with their critical 
counterparts.
We stress that scaling in the surface region is anisotropic, both 
in the critical and tricritical fixed points the correlation lengths 
are related as $\xi_{\parallel} \sim \xi_{\perp}^z$, with the same 
anisotropy exponent $z=2$. 
On the other hand the $x_s$ and $\beta_s$ exponents are different 
in the two cases, but the scaling relation $\beta_s=\nu_{\perp} x_s$ 
is satisfied in both fixed points. Notice also that the exponents in 
Table \ref{TAB03} are the same as for the restricted solid-on-solid 
model of the interface localization-delocalization transition 
\cite{lipowskykrollzia}.

\subsubsection{Crossover regime}

It is interesting to study the behavior of the surface magnetization in
the vicinity of the tricritical point when $0 < 1 - h_s \ll 0$.
This is a crossover region where the surface critical behavior for
finite $L$ is strongly influenced by the presence of the tricritical 
transition occurring at $h_s = 1$.

Analyzing the exact relations in (\ref{magnet1}-\ref{ki}) one can 
define
a length scale $\xi_s=1/2 |\ln h_s|$ 
which is identical to $\xi_{\perp}$ defined above in the first order 
case $h_s > 1$ in (\ref{xiperp}). Then one can show
that the finite size surface magnetization at the transition point, $t=0$,
satisfies the scaling relation:
\beqn
m_s(L,\ln h_s) & = & L^{-x_s^{\rm tr}} {\tilde m}_s(L/\xi_s)
\label{msscaling}
\eeqn
with $\xi_s$ defined above and where the scaling function 
${\tilde m}_s(y)$ goes to a constant for $y \to 0$ and 
behaves as ${\tilde m}_s(y) \sim 1/y^2$ for $y \to \infty$.
This scaling function is expected to be universal, provided 
the surface critical behavior of the model is independent 
of $Q>4$. We shall 
study this point numerically in the next Section.

\subsection{DMRG study}

\begin{figure}[b]
\centerline{\psfig{file=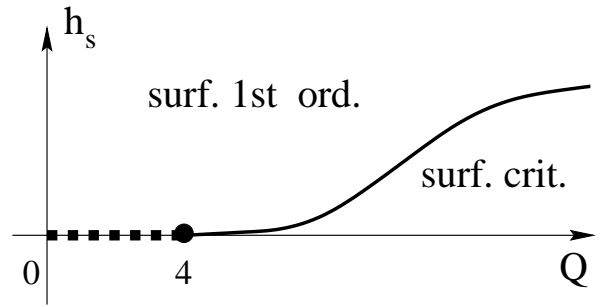,height=4cm}}
\vspace{5mm}
\caption{Schematic phase diagram for the surface critical behavior of
the Q-state Potts model as function of the applied surface 
field $h_s$. The thick solid line departing from $Q = 4$ 
denotes a tricritical line, which separates a continuous from 
a first order surface transition. For $Q \le 4$ (along the thick dashed
line) the surface critical exponents depend continuously on Q.}
\label{FIG04}
\end{figure}

Having the complete analytical solution for $Q \to \infty$ at 
hand, we are interested in the surface critical behavior for 
finite values of $Q$. The expected surface phase diagram of 
the model is schematically drawn in Fig. \ref{FIG04}.

In the second-order bulk transition regime, $Q \leq 4$, the 
critical and tricritical surface transition lines merge and 
are characterized by $Q$-dependent exponents. For $Q > 4$, 
in the first-order bulk transition regime, the second- and 
first-order surface transition regions are separated by a 
tricritical line, which is expected to depart from $Q=4$. 
It should also be noticed that some of the tricritical 
exponents obtained from the exact solution at $Q \to \infty$ 
are the same as for the $Q=4$ Potts model: $x_s^{tr}=x_s(Q=4)$ 
and $\nu_{\parallel}^{tr}=\nu_{\rm bulk}(Q=4)$, whereas other 
are different.
Then one can expect two scenarios about the surface (critical 
and tricritical) exponents of the $Q>4$ model: i) the exponents 
vary with $Q$, like for $Q \le 4$; or ii) the exponents are 
universal and their fixed-point values are those calculated 
at $Q \to \infty$.

To decide between the two possibilities
we have investigated the problem numerically using the density matrix
renormalization group. As for the bulk transition in Section III
we have considered the isotropic model for $5 \leq Q \leq 9$ on 
moderately large finite strips ($L \leq 40$) where we have calculated 
the magnetization profiles with fixed-free boundary conditions, using
the finite system algorithm described in \cite{whitePRB}. Comparing
the numerical data for different number of states kept in the DMRG 
procedure \cite{whitePRL} we could estimate the accuracy for the 
finite surface magnetization as about $5-6$ digits.

\begin{figure}[b]
\centerline{\psfig{file=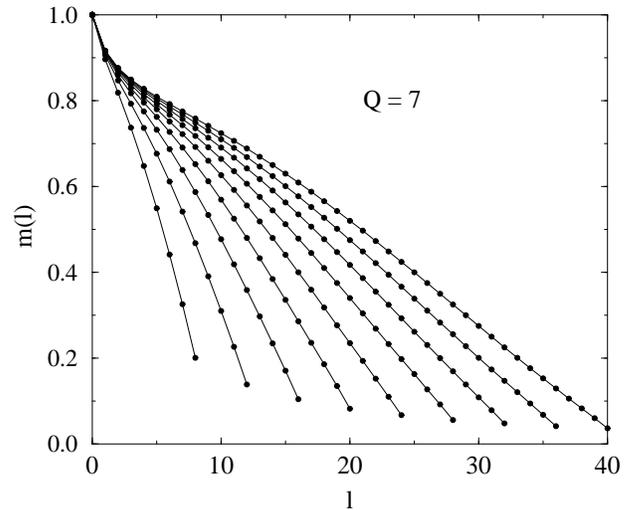,height=7cm}}
\vspace{5mm}
\caption{Magnetization profiles at the bulk transition point for
fixed/free boundary conditions and for $L=8$, $12$, $16$, \ldots,
$40$ with $Q=7$.}
\label{FIGQ5}
\end{figure}

\begin{figure}[b]
\centerline{\psfig{file=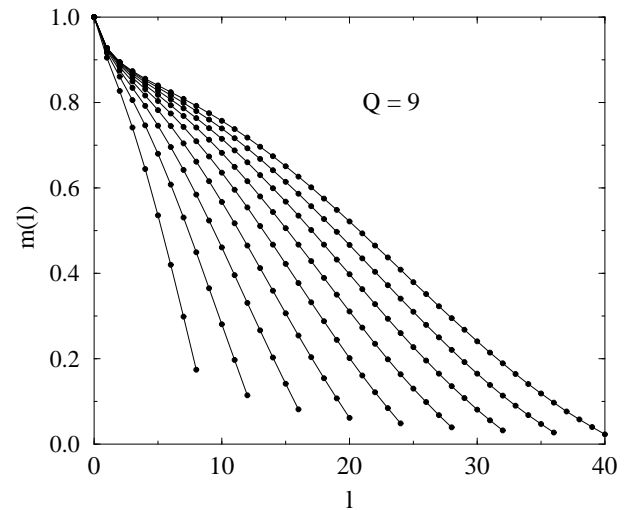,height=7cm}}
\vspace{5mm}
\caption{As in Fig. {\protect{\ref{FIGQ5}}} but for $Q=9$.}
\label{FIGQ6}
\end{figure}

Figures \ref{FIGQ5} and \ref{FIGQ6} show plots of magnetization 
profiles for the $Q = 7$ and $Q = 9$ cases, for different values 
of the strip width $L$.
From the surface magnetization $m_s(L)$ at the bulk transition point and
at $h_s=0$ we 
first formed finite-size extrapolants: $x_s(L)=\ln(m_s(L-4)/m_s(L))/
\ln(1-4/L)$ from the data for $L = 8$, $12$, $16$, \ldots $40$.
For $L \to \infty$ one has $x_s (L) \to x_s$. Table \ref{TAB04} 
shows the values of $x_s (L)$ for the cases $Q = 8$ and $Q = 9$. All 
the finite size data are above the value $x_s = 1$ (the surface 
critical dimension for the $Q = 4$ model) and tend to increase
at larger $Q$. Extrapolations from the BST sequence extrapolation 
method \cite{BST} yield $x_s=2.7(4)$ for $Q = 8$ and $x_s=2.9(3)$ 
for $Q = 9$,
both compatible with the value of $x_s = 3$ found in the $Q \to
\infty$ limit.
These estimates have quite large error bars (from this 
reason we could not get a sensitive estimate for $Q\le 7$) and we 
did not find nice convergence of the BST extrapolants for $x_s(L)$ as 
was found in the cases $Q=2$ and $Q=3$ in Ref.\onlinecite{carlonigloi}. 
This is typically due to strong finite size corrections, which in the 
present case are essentially caused by two effects.
First of all the bulk correlation length $\xi_b (Q)$, which remains
finite at a first-order transition is still very large for the values 
of $Q$ analyzed. As seen in the values of the bulk correlation lengths
in Table \ref{TAB02} only the highest values of $Q$
are reliable for extrapolations of finite size data with $L \leq 40$.
A second point is that in the range $5 \leq Q \leq 9$ we expect that 
the tricritical surface transition is at small surface fields, so that 
even at $h_s = 0$ strong influence from the tricritical fixed-point can 
be observed.

To clarify this point we have reanalyzed our numerical data using the 
analytical scaling relation in (\ref{msscaling}) for the $Q \to \infty$ 
limit, which describes the behavior of the surface magnetization at and 
in the vicinity of the tricritical point.
As already mentioned the scaling function in (\ref{msscaling}) 
is expected to be universal, provided the surface critical behavior of 
the model is independent of $Q>4$.
In the analysis of numerical results we have compared the ratio of 
surface magnetizations for a finite $Q$ and $h_s=0$, with the 
corresponding analytical results at $Q \to \infty$ and at finite 
$h_s$.
We found the value of $h_s$ such that the equality:
$m_s^Q(L)/m_s^Q(L-4)= m_s^{\infty}(L,h_s)/m_s^{\infty}(L-4,h_s)$
is satisfied. 
The universality hypothesis about the scaling
function is connected with the convergence of the finite size estimates
$h_s(L)$ to a finite limit. Moreover, the difference, $\Delta h_s=1-h_s$,
gives an estimate on the position of the tricritical line at a given
$Q>4$.

As seen in Table \ref{TAB04} the $\Delta h_s(L)$ finite size 
extrapolants for the $Q=7$ model are nicely converging to a 
limiting value and the same observation has been found for 
the other $Q$-s, as well. The extrapolated position of the 
tricritical line is found as $\Delta h_s=0.0125$, $0.0215$, 
$0.0310$ and $0.040$, for $Q=6,7,8$ and $9$, respectively \cite{note}. 
This is a strong numerical indication that the surface critical 
phenomena of the $Q>4$ state Potts model is universal and the 
corresponding exponents are those given in Table \ref{TAB03}.
The influence of the tricritical point is also visible in the 
profiles for $Q=7$ of Fig. \ref{FIGQ5} where the magnetization
has an approximately linear behavior as function of the distance from the free 
surface as found for the tricritical profile in the limit
$Q \to \infty$ (see dotted line of Fig. \ref{FIG03}). In the 
case $Q = 9$ instead, the curvature of the profile is clearly 
noticeable for the largest systems analyzed.

Finally, we turn to compare the two finite length scales appearing
in the problem:
the $\xi_s \approx 1/2\Delta h_s$ surface length scale and the $\xi_b$
bulk correlation length. 
Assuming anisotropic scaling as $\xi_b \sim \xi_s^{\theta}$, we obtain
$\theta=1.15,~1.24$ and $1.4$ for $Q=8,7$ and $6$, respectively. In the
$Q \to 4^+$ limit, when the correlation lengths are divergent,
it is natural to assume that the anisotropy exponent will approach
$\theta=2$, so that the
perpendicular and parallel correlation lengths are related as in the
critical and tricritical surface transitions (see Table \ref{TAB03}). 
With this assumption we can
conjecture the position of the tricritical line as
\be
\lim_{Q \to 4^+} \Delta h_s(Q) \sim \left[\xi_b(Q)\right]^{-1/2}
\sim \exp \left[-{\pi^2 \over 2 \sqrt{Q-4}}\right]\;,
\ee
where the limiting form of $\xi_b(Q)$ can be found in 
\cite{ksz}.

\section{Discussion}
\label{sec:disc}

In this paper the two-dimensional $Q$-state Potts model has been studied
by analytical and numerical methods in the first-order bulk transition
regime, $Q>4$. In the bulk transition point we have used the DMRG method
and sequence extrapolation techniques to accurately estimate the latent
heat and the magnetization discontinuity at the transition point. Calculating
the correlation lengths at the transition point we have shown with good
numerical accuracy that they are identical in the ordered and in the disordered phases.

For the surface transition we have demonstrated the existence of
surface induced disorder, for the first time for a microscopic 
model. We have calculated the complete set of critical and 
tricritical surface exponents and showed that the surface transition 
of the model is universal for $Q>4$, thus do not depend on the value 
of the discontinuities at the bulk transition point.
Based on this observation we conjecture that the surface exponents 
in Table \ref{TAB03} are universal for all two dimensional systems 
with a first-order bulk transition. This statement is in accordance 
with the prediction of the restricted solid-on-solid model of the 
interface localization-delocalization transition \cite{lipowskykrollzia}. 
It is interesting to note that the transfer matrix of the restricted
solid-on-solid model is formally equivalent to the Hamiltonian 
$\tilde{H}$ in (\ref{htilde}), which represents the ground-state block of the
anisotropic version of the $Q \to \infty$ Potts model. The physical
interpretation of the same operator in the two problems is, however, 
completely different.

\acknowledgements
F.I.'s work has been supported by the Hungarian National Research Fund under
grants No OTKA TO23642 and OTKA TO25139 and by the Ministry
of Education under grant No FKFP 0765/1997.
E.C. is financially supported by the Fund for Scientific Research of Flanders
(FWO G.0239.96).
Useful discussions with J. S\'olyom and L. Turban are gratefully acknowledged.

\appendix
\section*{Solution of the ground-state sector of ${\cal H}$}

Here we show how to construct the ground-state sector of (\ref{hamilton})
for the smallest systems, $L=2$ and $L=3$ and then use these results to
define a systematic $1/Q$-expansion in the vicinity of the bulk transition 
point.

The ground-state sector for the $L=2$ system is spanned by two vectors:
\beqn
\psi_1&=&11 \nonumber \\
\psi_2&=&{1 \over \sqrt{Q-1}}(12+13+\dots +1q) \nonumber \\
\label{L2}
\eeqn
and the eigenvalue matrix with zero surface field, $H_s=0$, is given by:
\be
{\tilde H^2} =\left(
\matrix{
 -1 & -h\sqrt{Q-1} \cr
 -h\sqrt{Q-1} & -h(Q-2) \cr}
\right)\;.
\label{h2}
\ee
For $L=3$ the ground-state sector is spanned by five vectors:
\beqn
\psi_1&=&111 \nonumber \\
\psi_2&=&{1 \over \sqrt{Q-1}}(112+113+\dots +11q) \nonumber \\
\psi_3&=&{1 \over \sqrt{Q-1}}(121+131+\dots +1q1) \nonumber \\
\psi_4&=&{1 \over \sqrt{Q-1}}(122+133+\dots +1qq) \nonumber \\
\psi_5&=&{1 \over \sqrt{(Q-1)(Q-2)}}(123+124+\dots+12q
\nonumber \\
&+& 132\dots +1q(q-1)) 
\label{L3}
\eeqn
and the eigenvalue matrix for $H_s=0$ is given by:
\end{multicols}\widetext
\be
{\tilde H^3} = \left(
\matrix{
-2 &  -h\sqrt{Q-1}  & -h\sqrt{Q-1}   &  0  &     0   \cr
 -h\sqrt{Q-1} &  -1-h(Q-2)  &   0  &  -h   & -h\sqrt{Q-2} \cr
-h\sqrt{Q-1}  &  0  & -h(Q-2)   & -h & -h\sqrt{Q-2}    \cr
 0  & -h  & -h & -1 &  -h2\sqrt{Q-2}  \cr
  0 & -h\sqrt{Q-2}  & -h\sqrt{Q-2} & -h2\sqrt{Q-2}   & -h2(Q-3) \cr}
\right)\;.
\label{h3}
\ee
\begin{multicols}{2} \narrowtext
It is remarkable that the dimension of the ground-state sector is independent
of the value of $Q$ and only depends on $L$. In the vicinity of the bulk transition
point $h_c=1/Q$ further simplifications take place in the $Q \to \infty$
limit. To see this, first, we consider the $L=2$ problem, where the two
vectors $\psi_1$ and $\psi_2$ are now degenerate with the same diagonal
matrix-elements, whereas the off-diagonal matrix-elements are $-h\sqrt{Q}$
in leading order of $1/Q$. Similarly, in the $L=3$ system there are three
asymptotically degenerate states $\psi_1$, $\psi_2$ and $\psi_5$ in
(\ref{L3}) with diagonal matrix-elements 
$\tilde H^3_{11}=\tilde H^3_{22}=\tilde H^3_{55}=-2$ and these states are
separated from the others by an amount of $O(1)$. Since the off-diagonal matrix-elements
between the relevant states are 
$\tilde H^3_{12}=\tilde H^3_{21}=\tilde H^3_{25}=\tilde H^3_{52}=-h \sqrt{Q}
\approx 1/\sqrt{Q}$ the two non-relevant states do not influence the behavior of
the ground state (and the first excited state) in the large $Q$ limit.

Generalizing the observation found in the $L=2$ and $L=3$ systems one can
show that in the vicinity of the bulk transition point in the $Q \to \infty$ limit
for general $L$ the ground state sector is spanned by $L$ asymptotically
degenerate vectors:
\end{multicols}\widetext
\beqn
\psi_1&=&11 \dots 1 \nonumber \\
\psi_2&=&{1 \over Q^{1/2}}(11 \dots 12+11 \dots 13+\dots +11 \dots 1Q) \nonumber \\
\psi_3&=&{1 \over Q}(11\dots123+11 \dots 24+\dots +11 \dots Q(Q-1)) \nonumber \\
\psi_4&=&{1 \over Q^{3/2}}(11\dots 1234+11\dots 1235+\dots +11 \dots 1Q(Q-1)(Q-2)) \nonumber \\
&\vdots& \nonumber \\
\psi_L&=&{1 \over Q^{(L-1)/2}}(1234 \dots L+\dots +Q(Q-1)\dots (Q-L)) \nonumber \\
\label{LL}
\eeqn
and the corresponding eigenvalue matrix is given in (\ref{htilde}).
\begin{multicols}{2} \narrowtext

\end{multicols}\widetext
\begin{table}[hb]
\vspace{5mm}
\caption{Table of the BST extrapolants for the latent heat
for the $Q=8$ Potts model. In the first column are the DMRG data for finite
systems with $L=8,12,16,\dots,40$.
(The parameter of the BST algorithm is $\epsilon=2$.)}
\label{TAB01}
\begin{tabular}{ccccccccc}
\vspace{.5mm}\\
0.335380 & 0.288286 & 0.312288 & 0.236212 & 0.242762 & 
0.242690 & 0.242082 & 0.241738 & 0.241674 \\
0.307476 & 0.272438 & 0.227422 & 0.243086 & 0.242684 & 
0.242082 & 0.241738 & 0.241674 &\\  
0.291096 & 0.263010 & 0.240970 & 0.242728 & 0.242084 & 
0.241736 & 0.241674 & &\\ 
0.280320 & 0.256912 & 0.242214 & 0.242068 & 0.241752 & 
0.241670 & & & \\ 
0.272726 & 0.252760 & 0.242126 & 0.241644 & 0.241686 & & & & \\
0.267128 & 0.249838 & 0.241868 & 0.241434 & & & & & \\ 
0.262864 & 0.247734 & 0.241658 & & & & & & \\
0.259538 & 0.246198 & & & & & & & \\
0.256894 & & & & & & & & \\
\vspace{.5mm}\\
\end{tabular}
\end{table}
\begin{multicols}{2} \narrowtext

\begin{table}[hb]
\caption{Physical quantities at the bulk transition point of the Q state
Potts model as found from finite size BST extrapolation of DMRG
data ($^*$ denotes the exact values).}
\label{TAB02}
\begin{tabular}{ccccccc}
\vspace{1mm}\\
$Q$&$\Delta E$& $\Delta E^{(*)}$&$\Delta m$&$\Delta m^{(*)}$
&$\xi_o$&$\xi_d^{(*)}$ \\
\vspace{1mm}\\
\tableline
\vspace{1mm}
5 & 0.028(3) & 0.0265 & 0.49(1) & 0.4921 & -- & 2512.  \\
\vspace{.5mm}
6 & 0.102(3) & 0.1007 & 0.66(1) & 0.6652 & -- & 158.9  \\
\vspace{.5mm}
8 & 0.242(2) & 0.2432 & 0.798(3) & 0.7998 & -- & 23.9  \\
\vspace{.5mm}
9 & 0.299(2) & 0.2998 & 0.833(2) & 0.8333 & 15.0(4) & 14.9  \\
\end{tabular}
\end{table}

\begin{table}[hb]
\caption{Surface critical and tricritical exponents of the Q state
Potts model calculated in the $Q \to \infty$ limit.}
\label{TAB03}
\begin{tabular}{ccc}
\vspace{1mm}\\
& critical & tricritical \\
\vspace{1mm}\\
\tableline
\vspace{1mm}
$x_s$ & 3 & 1  \\
\vspace{.5mm}
$\beta_s$& 1 & 1/3  \\
\vspace{.5mm}
$\nu_{\perp}$ &  1/3 & 1/3  \\
\vspace{.5mm}
$\nu_{\parallel}$& 2/3 & 2/3  \\
\vspace{.5mm}
$z$ & 2 & 2  \\
\end{tabular}
\end{table}

\begin{table}[hb]
\caption{Finite size extrapolants for the surface magnetization scaling 
dimension, $x_s^Q (L)$, and for the position of the tricritical line, 
$\Delta h_s^Q (L)$, as a function of the width of the strip for various 
values of $Q$. The last row shows the BST extrapolations at $L \to \infty$.
}
\label{TAB04}
\begin{tabular}{cccc}
\vspace{1mm}\\
L& $x_s^{Q=8}(L)$ & $x_s^{Q=9}(L)$ & 
$\Delta h_s^{Q=7}(L)$ \\
\vspace{1mm}\\
\tableline
\vspace{1mm}
20 & 1.15892328 & 1.25272509 & 0.013300 \\
\vspace{.5mm}
24 & 1.22149616 & 1.32785249 & 0.017501 \\
\vspace{.5mm}
28 & 1.27464891 & 1.39259493 & 0.019548 \\
\vspace{.5mm}
32 & 1.32114750 & 1.44961396 & 0.020599 \\
\vspace{.5mm}
36 & 1.36226612 & 1.50098953 & 0.021121 \\
\vspace{.5mm}
40 & 1.39920082 & 1.54783012 & 0.021344 \\
$\infty$ & 2.7(4) & 2.9(3) & 0.0215(1)
\end{tabular}
\end{table}

\end{multicols}

\begin{references}
\bibitem{note} For $Q=5$ the finite size effects,
due to the large bulk correlation length ($L/\xi_b \le 0.015$),
are still very strong;
the tricritical line is so close
to the $h_s = 0$ line 
that we could not extract its position from the available finite-size
data.
\end{references}
\end{document}